\documentclass{article}%
\usepackage{amsfonts}
\usepackage{amsmath}
\usepackage{amssymb}
\usepackage{graphicx}%
\begin{document}

\title{Condensate Nuclei and Magnetic Polarity Reversals in the Sun and Solar-type Stars}
\author{Babur M. Mirza\\Department of Mathematics, \\Quaid-i-Azam University, 45320, Islamabad. PK.}
\maketitle

\begin{abstract}
Magnetic field generation in the Sun and solar-type stars is modelled here
based on the formation of magnetically polarized condensates (B. M. Mirza,
\textit{Mod. Phys. Lett. B} \textbf{28} (2014) 1450148) under the high density
and pressure conditions. The correct orders of magnitude for the time period
and the energy loss in a solar cycle are deduced, as well as the enhancement
in the energy emission observed during solar cycles. It is shown that this
feedback magnetic field along with differential rotation is sufficient to
generate the toroidal magnetic field in the solar exterior. The model is
useful in determining magnetic polarity reversals, energy generation, and
reversal times, in solar-type stars.

\end{abstract}

\section{Introduction}

Magnetic field in normal stars like the Sun plays an important role almost all
stellar activity. Magnetic field governs many processes on the Sun including
the sunspot generation, particle acceleration effects causing solar
prominences and solar wind, and numerous other aspects of the solar corona,
the chromosphere and the photosphere [2]. In fact the Sun possesses a
large-scale poloidal field that is particularly prominent in the polar regions
and has a dipole symmetry. This magnetic field reverses polarity in a cycle of
about $11$ years. In general this cyclic activity is observed in most Sun-like
stars with periods around $10-11$ years [3].

On the solar surface, the deformation of toroidal field lines due to the
differential rotation of the Sun accounts for the sunspot cyclic behavior
associated with the solar cycle, particularly the pattern of sunspots around
the equator (Sp\"{o}rer's law). There are various processes by which this is
explained, including the formation and motion of magnetically active zones
called the convection cells ([4-6], also for further references see Ref. [7]).

However, the continuous generation of the magnetic field requires a feedback
mechanism. Observationally, this field is produced in the stellar core. In
fact almost all stellar energy, including the high poloidal magnetic field
(about $10G-10^{3}G$) energy of these stars, is generated in the stellar
nucleus. Also, helioseimic measurements for instance, indicate extreme
pressure and density conditions in the solar core. These conditions are
sufficient to generate mass-energy conversion in the solar interior [2, 8].

Here, it is shown that condensate formation is an efficient mechanism for
magnetic polarity reversals and energy generation in the Sun and other
solar-type stars. The model is based on the elementary theory of magnetically
polarized condensates.

The general plan of the work is as follows. Starting from the condensate
energy equation (detailed derivation can be found in Ref. [1]), we first
calculate (Section 2) the pressure required to counter-balance the
gravitational pressure at the condensate surface. This implies the equilibrium
condition under which a condensate can exist in the stellar core. This in turn
yields a harmonic constraint on the condensate radius, whose amplitude and
frequency are determined using quantum energy, and relativistic velocity
conditions. We then calculate (in Section 3) the energy loss in the process.
It is seen that the energy loss is maximum at each solar cycle periodically.
The question of how this condensate energy is converted into the magnetic
energy of the star is then discussed in Section 4. Here the electon-ion
magnetic polarity is used to assign a micro magnetic field via the molecular
magnetic field coefficient to the particles forming the condensate. It gives
absolute magnetization as a measure for the magnetic field strength of the
star. Based on this model, and incorporating the mass-energy conversion
correction, correct order of magnitude for magnetic field reversal times
(roughly equal to the $10-11$ year solar cycle) is obtained. Futhermore, the
magnetic field strength at the time of field reversals, shows a marked
increase in magnitude, as observed during solar cycles. Section 5 gives a
summary of the main results.

\section{Condensate Pressure and the Equilibrium Condition}

In normal main sequence stars, and generally for gravitationally bound
systems, a stable equilibrium configuration exists when internal thermal
pressure is counterbalanced by the gravitational pressure. This thermal
pressure is primarily due to the mean kinetic energy of the ensemble under the
ideal gas assumption. Inside the stellar core high density conditions force
particles to coalesce, especially in the limiting condition for matter close
to the center of the star.

Thermodynamically the pressure set up at any point in the star is given by the
relation:%
\begin{equation}
p=-\left(  \frac{\partial\epsilon}{\partial V}\right)  _{S}, \tag{1}%
\end{equation}
where $S$ is the entropy, and $\epsilon$ is the macroscopic energy of the
system. For a condensate this energy can be calculated using the expression :%
\begin{equation}
\epsilon=\frac{2\pi\hbar}{\bar{\lambda}}\bar{v}\tan\left[  2\pi\left(
\frac{\bigtriangleup x-\bar{v}\bigtriangleup t}{\bar{\lambda}}\right)
\right]  . \tag{2}%
\end{equation}
(see Ref. [1] for the derivation) where $\bar{v}$ is the average particle
speed and $\bar{\lambda}$ is the average particles wavelength for the system.
Let the system has volume $V$ containing $N$ particles, then each linear
dimension of this volume can be given as $\bigtriangleup x\approx\left(
2^{3}V/N\right)  ^{1/3}$. Thus the thermal (total) energy of the condensate is
given by%
\begin{equation}
\epsilon=h\nu\tan\left[  \frac{2\pi}{\bar{\lambda}}\left(  \frac{2^{3}V}%
{N}\right)  ^{1/3}-2\pi\bar{\nu}t\right]  . \tag{3}%
\end{equation}
In equation (3) we have chosen $t_{0}=0$, thus we have substituted
$\bigtriangleup t=t$. Also $\bar{\nu}=\bar{v}/\bar{\lambda}$ is the (average)
particle related frequency. Using equations (1) and (3) we find that the
pressure for the condensate is given by:%
\begin{equation}
p=-\frac{\eta}{V^{2/3}}\sec^{2}\left[  \frac{2\pi}{\bar{\lambda}}\left(
\frac{2^{3}V}{N}\right)  ^{1/3}-2\pi\bar{\nu}t\right]  , \tag{4}%
\end{equation}
where $\eta=4\pi h\bar{v}/3\bar{\lambda}^{2}N^{1/3}$. This is the condensate
pressure. We notice that it depends on the volume and number density of the
particles in the condensate, and it gains a particularly large value when the
condensate radius becomes small, such as due to gravitational contraction.

On the other hand, for a spherically symmetric mass distribution the
gravitational pressure inside the star, at a distance $R$ from the center is%
\begin{equation}
p_{g}=\frac{3}{8\pi}\frac{GM^{2}}{R^{4}}, \tag{5}%
\end{equation}
where $M=M(R)$ is the total mass contained inside the sphere of radius $R$.

Now at each point of the condensate, at a distance $R$ from the center, the
outward condensate pressure (4) must equal the inward gravitational pressure
$p_{g}$ given by equation (5). Thus the equilibrium condition $p=-p_{g}$ for
the condensate, say at the condensate surface $R$, imply that:%

\begin{equation}
\frac{3GM^{2}}{8\pi\eta}\cos^{2}\left[  2\pi\left(  \frac{\bigtriangleup
x}{\bar{\lambda}}-2\pi\bar{\nu}t\right)  \right]  =R^{4}V^{-2/3}. \tag{6}%
\end{equation}
Since under the conditions the condensate has a spherical symmetry, therefore
we take for its volume $V=4\pi R^{3}/3$. This gives $R^{4}V^{-2/3}=4\pi
R^{2}/3$ and equation (6) gives the condensate radius:%

\begin{equation}
R(t)=\sqrt{\frac{9G\tilde{M}^{2}}{32\pi^{2}\eta}}\cos\left[  2\pi\left(
\frac{\bigtriangleup x}{\bar{\lambda}}-2\pi\bar{\nu}t\right)  \right]  .
\tag{7}%
\end{equation}
Equation (7) shows that the condensate radius, hence its volume, do not remain
constant but periodically oscillate from maximum radius $R_{\max}%
=\sqrt{9GM^{2}/32\pi^{2}\eta}$ to zero. Also note that the condensate mass
$\tilde{M}$ here is small compared to the mass of the solar nuclei.

We next calculate the frequency and time period of this oscillation for a
Sun-like star. First note that the periodicity of these oscillations depends
on the average particle frequency $\bar{\nu}$. In general this frequency is
given by the quantum condition $E=h\nu$. In a degenerate gas each electron/ion
is trapped in a fixed region of space, say a box of length $a$, with energy%
\begin{equation}
E=\frac{h^{2}}{8ma^{2}}. \tag{8}%
\end{equation}
Since it is the gravitational pressure of the surrounding gas that confines
the electron within this region, therefore its kinetic energy is given by
\begin{equation}
\frac{mv^{2}}{2}=\frac{GMm}{a}. \tag{9}%
\end{equation}
Notice that as the stellar core contracts the particle speed $v$ can be
exceedingly large. In the limiting case this velocity has an upper bound given
by the speed of light $c$. Thus combining equation (8) and (9), and using the
quantum energy law $E=h\nu$, the particle related frequency in the condensate
is given by $hc^{2}/16GMm$. This implies a fundamental mode of oscillations
with frequency:%
\begin{equation}
\Omega_{C}=\frac{h}{8m}\left(  \frac{c^{2}}{2GM}\right)  ^{2}, \tag{10}%
\end{equation}
for the star, which dependents only on the stellar mass $M$ and the mass of
the condensate particles. Putting in the for the parameters $M$ $_{\odot
}\approx2\times10^{30}kg$, $G=$ $6.673\times10^{-11}Nm/kg^{2}$, $m=9.109\times
10^{-31}kg$, $c=3\times10^{8}m/\sec$, and $h=6.626\times10^{-34}J\sec$, we
obtain for the solar condensate frequency:%
\begin{equation}
\Omega_{C\odot}\approx0\cdot1\times10^{-10}cycles/\sec. \tag{11}%
\end{equation}
Correspondingly the time to complete one such cycle is%
\begin{equation}
\tau_{C\odot}\approx10^{11}\sec. \tag{12}%
\end{equation}
This value differs by about $340$ times from the solar cycle period of
$11yrs\approx0.31536\times10^{9}\sec$. This difference in the values is due to
mass-energy conversion in the stellar interiors. The relativistic correction
to the time $\tau_{C\odot}$ (time-dilation) is of the order $10^{-2}\sec$ for
Sun. The calculated time of a solar cycle thus lies in the observed range of magnitudes.

\section{Energy Loss and the Role of Thermal Pressure}

An important aspect of the solar cycle is the enhanced energy generation
during the process. In this respect condensate squeezing by the gravitational
pressure acts as an effective mechanism for energy extraction from
gravitationally bound material systems. This process occurs at long time
scales, as indicated by the formula (10) for the Sun-like stars.

Using the energy formula (2), we find that the energy loss $-d\epsilon/dt$
during a solar cycle is given by%
\begin{equation}
-\frac{d\epsilon}{dt}=2\pi\nu E_{0}\sec^{2}\left[  \frac{2\pi\bigtriangleup
x}{\bar{\lambda}}-2\pi\nu t\right]  , \tag{13}%
\end{equation}
where $E_{0}=h\nu$ is the ground state (or minimum) energy of the bound
quantized system. This energy loss periodically reaches a maximum after a
time:
\begin{equation}
\tau_{C}=\frac{16G^{2}M^{2}m}{hc^{4}}. \tag{14}%
\end{equation}
The plot in Figure1 shows that the energy loss has an almost saturated value
except at the states of cyclic periodicity, where it is particularly high.

Finally, we estimate the temperature attained in the stellar nuclei in this
process. This can be derived by using the fact that the thermal equilibrium is
maintained by the counterbalancing of the thermal and gravitational pressure
(given by equation (5)), where as the thermal pressure is given by%

\begin{equation}
p_{T}=nk_{B}T, \tag{15}%
\end{equation}
where $n=3M/4\pi R^{3}m_{M}$. Thus the temperature supplied to the condensate
is:%
\begin{equation}
T=\frac{Gm_{H}}{2k_{B}}\frac{M(R)}{R} \tag{16}%
\end{equation}
This shows that as the stellar nuclei contracts $R\rightarrow0$, the
temperature becomes exceedingly large. The cyclic energy loss results in heat
generation in the stellar core, which in turn causes the general heating and
enhanced radiation emission in the stellar exterior during the solar cycle..

\section{Magnetic Polarity Reversals}

The energy generated in a solar cycle strongly effects the solar magnetic
field, in particular, it exhibits solar magnetic polarity reversals and an
enhancement in the solar magnetic intensity.

We note that the condensate must have a resultant magnetic polarity since it
is formed of particles possessing some net magnetic moment. This magnetic
polarity can therefore be calculated from the molecular field coefficient and
magnetic susceptibility of the system.

In general, the total energy $\epsilon$ and spontaneous magnetization $I$ of a
magnetized state of matter is related by
\begin{equation}
\epsilon=\epsilon_{1}-wI^{2}/2, \tag{17}%
\end{equation}
where $w$ is the molecular field coefficient, and $\epsilon_{1}$ is the
reference magnetic energy before magnetization. Also the magnetic
susceptibility of a material is given by $\chi=dI/dB$, where $B$ is the
magnetic field intensity. From these relations, and the constant $\epsilon
_{1}=0$, absolute magnetization can be expressed as
\begin{equation}
\mid I\mid=\sigma\tan^{1/2}\left[  \frac{2\pi}{\bar{\lambda}}\left(
\frac{2^{3}V}{N}\right)  ^{1/3}-2\pi\nu t\right]  , \tag{18}%
\end{equation}
where $\sigma$ denotes $\left(  2E_{0}/w\right)  ^{1/2}$, and where use has
been made of equation (3). Generally $\chi$ is not a function of $H$, so that
the magnetic intensity is related to magnetization by the equation $\chi B=\pm
I_{0}\pm I$, where the $\pm$ sign is due to the absolute value of $I$
appearing in equation (18). Thus we have for the magnitude of the magnetic
field%
\begin{equation}
\mid B\mid=\frac{\sigma}{\chi}\tan^{1/2}\left[  \frac{2\pi}{\bar{\lambda}%
}\left(  \frac{2^{3}V}{N}\right)  ^{1/3}-2\pi\nu t\right]  +B_{0} \tag{19}%
\end{equation}
where $B_{0}=\pm I_{0}/\chi$. Since the magnetic field becomes imaginary for
some values of $t$, to obtain the real values of the magnetic field $B$ we
multiply it with its complex conjugate, and then take the square-root. This is
essentially a dipolar field, since magnetization $I$ must has an orientation
in space. Also, since this field is generated in the stellar interior without
a rotational mechanism, it must be identified with the primary poloidal
magnetic field of the star. A plot of this magnetic field $B$ is given in
Fig.2, which shows a polarity reversal and a particularly enhanced magnetic
field intensity near after each cycle.

\section{Conclusions}

The above study leads to the following results:

(1) Under the extreme density and pressure, along with a high temperature,
maintained inside a Sun-like star, matter in normal stellar interiors
transforms into a condensate-like coherent state.

(2) This condensate however is not stabilized against the gravitational
pressure, which causes it to decay in size.

(3) The gravitational sqeeze-in of the condensate act as an active mechanism
of converting stellar matter into energy in the form of magnetic field, and
this occurs cyclically, resulting in the magnetic pole reversals and
heightened magnetization of the Sun.

(4) The condensate nucleus has a magnetic polarity and causes the generation
of the feedback poloidal magnetic field of the star. This poloidal field along
with the differential rotation of the star, causes the toroidal field
generation, and the observed cyclicity in the sunspot behavior.

Since condensate formation under high density conditions must be a rather
common feature in all gravitationally bound dense systems, it is of interest
to study its role in such systems as well, particularly in compact stars and
in the problem of gravitational stability (see Ref.[9]).

\textbf{Figure Captions:}

Figure 1: Energy loss due to condensate formation-decay oscillation per solar
cycle ($\approx0.31536\times10^{9}\sec$), for $\bigtriangleup x=4\bar{\lambda
}$. The net energy loss is divided here by $2\pi\nu E_{0}$ units, in equation (13).

Figure 2: Magnetic polarity reversals in a solar-type star ($M=M_{\odot}$),
with periodicity determined from equation (14) and relativistic correction of
factor $10^{-2}\sec$. Here $B_{0}=0$, $\bigtriangleup x=\bar{\lambda}$, and
$\sigma=-\chi$.

\section{}
\end{document}